\documentstyle[12pt,aps,epsf]{revtex}
\begin{document}
\baselineskip 0.3in
\title{Structures of Rotating Traditional Neutron Stars and Hyperon Stars in the Relativistic $\sigma-\omega$
Model}
\author{ De-hua Wen$^{1,2}$, Wei Chen$^{3}$, Xian-ju Wang$^{1}$, Bao-quan Ai$^{1}$ \\
 Guo-tao Liu$^{1}$, Dong-qiao Dong$^{1}$, Liang-gang Liu$^{1}$} \maketitle
\begin{center}
 $^{1}$Department of Physics, Zhongshan
University, Guangzhou 510275, China \\
$^{2}$Department of Physics, South China University of Technology,
Guangzhou 510641, China\\
$^{3}$Department of Physics, Jinan University, Guangzhou 510632, China\\
\end{center}
\setlength {\baselineskip} {20pt}

\begin{abstract}
\setlength {\baselineskip} {20pt}
 The influence of the rotation on the total masses and radii of the neutron stars are calculated by
the Hartle's slow rotation formalism, while the equation of state
is considered in a relativistic $\sigma-\omega$ model. Comparing
with the observation, the calculating result shows that the double
neutron star binaries are more like hyperon stars and the neutron
stars of X-ray binaries are more like traditional neutron stars.
As the changes of the mass and radius to a real neutron star
caused by the rotation are very small comparing with the total
mass and radius, one can see that Hartle's approximate method is
rational to deal with the rotating neutron stars. If three
property values: mass, radius and period are observed to the same
neutron star, then the EOS of this neutron star could be decided
entirely.
\end{abstract}

\section{Introduction}
\indent Neutron stars are dense, neutron-packed remnants of
massive stars after their supernova explosions. Recently, in both
experiment and theory, much concern is focused on the neutron
stars\cite{s1,s2,s3}. One reason is to determine the equation of
state (EOS) of superdense matter, and then understand the early
universe, its evolution to the present day and the various
astrophysical phenomena; the other one is that neutron star is one
of the more possible sources of detectable gravitational waves. To
understand neutron stars, the first thing is to understand its
structure, such as the compositions, total masses, radii,
redshifts etc. Because of the strongly gravitational field and the
high density, neutron stars must be studied in the framework of
general relativity. For a static neutron star, giving EOS, using
TOV equations\cite{s4} , which are educed from Einstein field
equations, one can get an exact solution. But for a rotating
neutron star, the components of Einstein field equations become
much difficult. Nowadays, several approximate solutions of this
problem were developed\cite{s5,s6}, in which Hartle's slow
rotation formalism\cite{s7,s8} is the most popular one.\\
\indent The properties of neutron stars such as masses, rotational
frequencies, radii, moments of inertia and redshifts are sensitive
to the EOS of the matters\cite{s2,s3}. As the interior core
contains most of the mass of neutron star, so attention to the EOS
is mostly focused on the core of neutron star, that is to the
matters at density above several times of nuclear matter
saturation density. In the core of neutron star, the compositions
still keep blurry in some degree due to the high density and
uncertainty of strong interaction, but as density increasing in
the neutron star, neutrons can drip out of nuclei and form a
neutron gas; and due to the URCA process $n\rightarrow
p+e^{-}+\bar{\nu}_{e}$ and the neutrality of neutron star, there
are at least protons and electrons in the chemical equilibrium; on
the other hand, as the density increasing, hyperons will be
dominant in the neutron stars\cite{s9}. In this letter, the
Hartle's formalism will be used to deal with two kinds of rotating
neutron stars: the traditional neutron stars, in which
$n,p,e,\mu,$ are the main elements; and hyperon stars, in which
$n,p,e,\mu,\Lambda,\Sigma,\Xi,\Delta$ are the main elements. The
EOSs of them will be considered in the relativistic
$\sigma-\omega$ model.\\
\indent In this letter, we adopt the metric signature - + + +,
G=c=1 (the factors G and c will be reinserted correctly in
numerical calculation).

\section{Hartle's slow rotation formalism}
\label{Hartle's slow rotation formalism} \indent In relativity,
the space-time geometry of a rotating star in equilibrium is
described by a stationary and axisymmetric metric of the form
\begin{equation}
ds^{2}=-e^{2\nu}dt^{2}+e^{2\lambda}dr^{2}+e^{2\psi}(d\phi-\omega
dt)^{2}+e^{2\mu}d\theta^{2}
\end{equation}
where $\omega(r)$ is the angular velocity of the local inertial
frame and is proportional to the star's rotational frequency
$\Omega$, which is the uniform angular velocity of the star
relative to an observer at infinity. Expanding the metric function
through second order in $\Omega$, one has\cite{s7}
\begin{equation}\label{2-a}
e^{2\nu}=e^{2\varphi}[1+2(h_{0}+h_{2}P_{2})],
\end{equation}
\begin{equation}\label{2-b}
e^{2\lambda}=[1+\frac{2}{r}(m_{0}+m_{2}P_{2})(1-\frac{2M_{0}(r)}{r})^{-1}](1-\frac{2M_{0}(r)}{r})^{-1},
\end{equation}
\begin{equation}\label{2-c}
e^{2\psi}=r^{2}sin^{2}\theta[1+2(v_{2}-h_{2})P_{2}],
\end{equation}
\begin{equation}\label{2-d}
e^{2\mu}=r^{2}[1+2(v_{2}-h_{2})P_{2}],
\end{equation}
where $e^{2\varphi}$ and $M_{0}(r)$  denote the metric function
and the mass of the nonrotaiting neutron star with the same
central density, respectively; $P_{2}$  is the Legendre ploynomial
of order 2; the perturbation functions
$m_{0},m_{2},h_{0},h_{2},v_{2}$ are proportional to $\Omega^{2}$
and are to be calculated from Einstein field equations.\\
\indent From the $(t,\phi)$ component of Einstein field equations,
one gets\cite{s7}
\begin{equation}
\frac{1}{r^{4}}\frac{d}{dr}(r^{4}j\frac{d\bar{\omega}}{dr})+\frac{4}{r}\frac{dj}{dr}\bar{\omega}=0,
\end{equation}
where$j(r)=e^{-\varphi}[1-2M_{0}(r)/r]^{\frac{1}{2}}$,
$\bar{\omega}=\Omega-\omega$, which denotes the angular velocity
of the fluid relative to the local inertial frame. The boundary
conditions are imposed as $\bar{\omega}=\bar{\omega}_{c}$ at the
center, $ \frac{d\bar{\omega}}{dr}|_{\bar{\omega}_{c}=0}$ , where
 $\bar{\omega}_{c}$ is chosen arbitrarily. Integrating eq.(6)
outward from the center of the star, one would get the function
$\bar{\omega}(r)$. Outside the star, from eq.(6) one has
$\bar{\omega}(r)=\Omega-\frac{2J}{r^{3}}$, where $J$ is the total
angular mementum of the star, which takes the form\cite{s2}
$J=\frac{1}{6}R^{4}_{0}\frac{d\bar{\omega}}{dr}|_{r=R_{0}}$. So at
the surface, one can determine the angular velocity $\Omega$
corresponding to $\bar{\omega}_{c}$ as
\begin{equation}\
\Omega=\bar{\omega}(R_{0})+2\frac{J}{R^{3}_{0}}.
\end{equation}

From the $(t,t)$  and $(r,r)$  components of Einstein field
equations, one gets two coupled ordinary differential equations of
$m_{0}$ and $h_{0}$ as\cite{s7,s8}
\begin{equation}\
\frac{dm_{0}}{dr}=4\pi
r^{2}\frac{d(p+\rho)}{dp}(\rho+p)p^{*}_{0}+\frac{1}{12}j^{2}r^{4}(\frac{d\bar{\omega}}{dr})^{2}-\frac{1}{3}
r^{3}\frac{d(j^{2})}{dr}\bar{\omega}^{2},
\end{equation}
\begin{equation}\
\frac{dp_{0}^{*}}{dr}=-\frac{m_{0}(1+8\pi
r^{2}p)}{[r-2M_{0}(r)]^{2}}-\frac{4\pi
r^{2}(p+\rho)}{r-2M_{0}(r)}p^{*}_{0}+\frac{1}{12}\frac{r^{4}j^{2}}{r-2M_{0}(r)}(\frac{d\bar{\omega}}{dr})^{2}
+\frac{1}{3}\frac{d}{dr}[\frac{r^{3}j^{2}\bar{\omega}^{2}}{r-2M_{0}(r)}],
\end{equation}
 where $p^{*}_{0}=-h_{0}+\frac{1}{3}r^{2}e^{-2\upsilon}\bar{\omega}^{2}+C$
, here $C$ is a constant determined by the demand that $h_{0}$ be
continuous across the star's surface. These equations are also
integrated outward, with boundary conditions that both $m_{0}$ and
$p^{*}_{0}$  vanish at the origin. With the same central density,
the difference between the mass of the rotating star and the
non-rotating star is
\begin{equation}\
\delta M=m_{0}(R_{0})+\frac{J^{2}}{R_{0}^{3}},
\end{equation}
The difference of the mean radius is
\begin{equation}\
\delta r=-p^{*}_{0}(\rho+p)\frac{dp}{dr}.
\end{equation}

\section{Model for the EOSs $-$ the relativistic $\sigma-\omega$ model}
\indent There are several models to deal with the superdense
matters, such as non-relativistic models, relativistic
fieldtheoretical models\cite{s3}. To different models, the
fracions of particles in the superdense matters are different, and
then the bulk properties of superdense matters are different, that
is, the EOS of them are different. Here the Relativistic
$\sigma-\omega$ Model will be adopted\cite{s10}. The Lagrangian
density of this model is
\begin{eqnarray}
{\it
L}&=&\sum_{B}\bar{\psi}_{B}(i\gamma_{\mu}\partial^{\mu}+m_{B}-g_{\sigma
B}\sigma-
g_{\omega B}\gamma_{\mu}\omega^{\mu}-\frac{1}{2}g_{\rho B}\gamma_{\mu}\vec{\tau}.\vec{\rho}^{\mu})\psi_{B}+\nonumber \\
&
&\frac{1}{2}(\partial\sigma)^{2}-\frac{1}{2}m_{\sigma}^{2}\sigma^{2}-
\frac{1}{4}F_{\mu\nu}F^{\mu\nu}+\frac{1}{2}m_{\omega}^{2}\omega_{\mu}\omega^{\mu}-U(\sigma)-\nonumber \\
& &\frac{1}{4}\vec{\rho}_{\mu\nu}.\vec{\rho}^{\mu\nu}+
\frac{1}{2}m_{\rho}^{2}\vec{\rho}_{\mu}.\vec{\rho}^{\mu}+
 \sum_{l}\bar{\psi}_{l}(i\gamma_{\mu}\partial^{\mu}-m_{l})\psi_{l}
\end{eqnarray}
in which
$U(\sigma)=a\sigma+\frac{1}{3!}c\sigma^{3}+\frac{1}{4!}d\sigma^{4}$,
$F_{\mu\nu}=\partial_{\mu}\omega_{\nu}-\partial_{\nu}\omega_{\mu}$,
$\vec{\rho}_{\mu\nu}=\partial_{\mu}\vec{\rho}_{\nu}-\partial_{\nu}\vec{\rho}_{\mu}$,
 $\psi_{B}$ is the field operator of Baryon $B$($B$=n, p for the traditional neutron stars, $B$=n, p,
$\Lambda$, $\Sigma$, $\Xi$, $\Delta$ for the hyperon stars);
$\psi_{l}$ is the field operator of lepton $l$($l$=$e$, $\mu$);
and $\sigma$, $\omega^{\mu}$, $\vec{\rho}^{\mu}$ are the field
operators of $\sigma$, $\omega$, $\rho$ meson respectively;
$g_{\sigma B}$, $g_{\omega B}$, $g_{\rho B}$ are the coupling
constants between $\sigma$, $\omega$, $\rho$ meson and baryon $B$
respectively. In general, the coupling constants between
$\sigma$($\omega$ or $\rho$) meson and neutron and proton are
equal and decided by saturated property of nuclear
matter($g_{\sigma n}=g_{\sigma p},g_{\omega n}=g_{\omega p}$) and
the symmetry energy of nuclear matter($g_{\sigma \rho}, g_{\omega
\rho}$); $m_{B},m_{l},m_{i},(i=\sigma,\omega,\rho)$ are the mass
of baryon, lepton, meson respectively; $\vec{\tau}$ is the isospin
operator. To leptons, we assume they are free fermi gas. From this
Laglangian density, we can obtain the EOS of the superdense
matters, $p=p(\rho)$, in which, $p$ and $\rho$ are the pressure
and energy density of the superdense matters, respectively. In
1-loop approximation, the loop's contribution to the propagators
of nucleon and $\sigma$ meson is considered, and the
renormalization is used to renormalizing the divergent part of
loop contribution. In the numerical calculation, we adopt the
value of the parameter as following\cite{s10}: $a=-2.1\times
10^{7}MeV^{3}$, $c=0.97\times M_{n}$, $d=1277$, $g_{s}=6.73$,
$g_{v}=8.59$,  $M_{n}=938MeV$, $m_{\omega}=783MeV$,
$m_{\sigma}=550MeV$, $m_{\rho}=770MeV$, and the incompressibility
of nuclear matter is $224MeV$, which is consistent with the
experiment result\cite{s11,s12}. From Fig.1, it is clear that the
EOS of traditional neutron stars is stiffer than the EOS of
hyperon stars.
\begin{center}
\textbf{Fig.1}
\end{center}

\section{Numerical results and discussion}
\indent In order to compare the numerical results with the
observation, we present some typical observation value here. As we
know, only a few masses have been determined in observation from
the more than thousand neutron stars. There are two typical
observed mass data: $1.36\pm0.08M_{\bigodot}$\cite{s13}  for
double neutron star binaries, and
$1.87^{+0.23}_{-0.17}M_{\bigodot} $(Vela X-1)\cite{s14} and
$1.8\pm0.4M_{\bigodot}$(Cygnus X-2)\cite{s15}
for X-ray binaries. \\
\indent The numerical results for the rotating traditional neutron
stars and hyperon stars in the relativistic $\sigma-\omega$ model
are showed in
the following figures and table.\\
 \indent Figs.2-3 show the changes
of masses and radii between rotating neutron stars and
non-rotating neutron stars as function of rotational period. At a
given central density, it is easy to see that, as the neutron
stars rotating slowly, the increment of the masses and radii will
reduce sharply when the neutron stars' rotational periods are
small than 1 ms. But as the periods become bigger than 1.6ms,
which is the period of the fastest rotating pulsars in
observation\cite{s3}, the change of the increment of the masses
and radii caused by the rotation will become very week, and the
increment will be no more than 2 percent. As the changes of the
mass and radius to a real neutron star caused by the rotation are
very small comparing with the total mass and radius, we can see
that Hartle's approximate method is rational.\\
\indent  Figs.4-7 show the masses and radii of rotating and
non-rotating traditional neutron stars and hyperon stars as
function of central density at a given central angular velocity
relative to the local inertial frame. From these figures, one can
see that around the typical observational radii with value of
12km, the increments of the radii are bigger than that of other
place. In table 1, some typical value in these figures are listed,
combining fig.4, fig.6 and the observational value, one can see
that the double neutron star binaries are more like hyperon stars,
while the neutron stars of X-ray binaries are more like
traditional neutron stars. For the hyperon stars, its EOS is so
soft that, even the period small than the smallest observational
value, the total mass can't reach the observation value of X-ray
binaries, about $1.8M_{\bigodot}$. \\
\indent Fig.8 shows the period as a function of the central
density at two giving central angular velocities relative to the
local inertial frame. From this figure, one can see that bigger
central density and bigger central angular velocity relative to
the local inertial frame correspond with a smaller period. In
fig.8 one can find out that as the EOS of hyperon star is softer,
at the same conditions, the period of hyperon star is appreciably
bigger than the period of the traditional neutron star. Figs.9-10
show the masses and radii of the traditional neutron stars and
hyperon stars as a function of period. One can see that even the
central density of the hyperon stars is bigger than that of the
traditional neutron stars, the masses and radii of the hyperon
stars are smaller than that of the traditional neutron stars.
Another interesting result is that at a giving central density,
when the period increases to a value which is bigger than the
smallest observational period, 1.6ms, the change of the masses and
the radii with the period is not obvious,
which is consistent with the result of fig.2. \\
\indent From these figures, one can see that at a giving period,
the central density and the central angular velocity relative to
the local inertial frame could be chosen freely. but if the masses
and the period of a neutron star are given (by the observational
value) at the same time, then the central density and the central
angular velocity relative to the local inertial frame will be
decided. As we know, there is another observational value: the
radius of the neutron star, the decided central density and
central angular are not always given the exactly radius. So one
can say that if these three observational values: mass, radius and
period are given to the same neutron star, there is only one
special EOS could give a set of calculating values, which are fit
like a glove to the set of observational value, that is, these
three observational values could decide the EOS entirely. But the
problem is that there is no neutron star with both mass and radius
observationally determined up to the present.\\
\begin{center} \textbf{Fig. 2-Fig. 10}
\end{center}
 \begin{center}
 Table 1.      Rotating neutron star's property in the
      relativistic $\sigma-\omega$ model$^{*}$
 \end{center}
\begin{tabular}{c|cccccccccc}
   \hline
   & $\bar{\omega}_{c}$  & $\rho_{c}$ & $R_{0}(km)$ & $M_{0}$ & $R(km)$
   & $M$ & $\frac{\delta R}{R_{0}}$&$\frac{\delta M}{M_{0}}$&$P(ms)$\\
  \hline
  TNS  &2.50 &1.067 &11.98 &1.768 &12.18 &1.808 &0.017 &0.023 &1.488\\
   &5.00 &0.903 &10.21 & 1.650 &12.94 &1.801 &0.070 &0.092 &0.791\\
\hline
HS&2.50 &0.921 &11.80 &1.336 &12.04 &1.367 &0.020 &0.023 &1.686\\
  &5.00 &0.811 &10.87 &1.265 &12.87 &1.385 &0.084 &0.095  &0.873\\
 \hline
\end{tabular}\\

{\footnotesize $^{*}$In the table, TNS and HS denote the
traditional neutron stars and the hyperon stars respectively;
$\bar{\omega}_{c}$ is the angular velocity relative to the local
inertial frame at the center, with a unit of $(10^{3}s^{-1})$;
$\rho_{c} $ is the central density, with a unit of
$(10^{18}kg.m^{-3})$; $R_{0} $ and $M_{0}$ denote the radius and
the mass of the non-rotating stars, $R $ and $M$
      denote the radius and the mass of the rotating neutron stars,
      respectively; the unit of the mass is the solar mass$
      (M_{\bigodot})$, $\frac{\delta R}{R_{0}}$ and $\frac{\delta M}{M_{0}}$ are the fractional difference
      of the radius and mass between the rotating neutron stars and non-rotating neutron stars;
       P is the rotational period.\\}

{\bf Acknowledgements}\\
 \indent The project supported by National
Natural Science Foundation of China (Grant No. of 10275099,
10175096),  GuangDong Provincial Natural Science Foundation (Grant
No. of 021707) and Natural Science Foundation of
SCUT (Grant No. of E5123266).\\

\section{Figure captions}
\baselineskip 0.3in

Fig. 1. Equation of state of traditional neutron stars(TNS) and
hyperon stars(HS).

Fig. 2. Increment of the mass and radius of TNS comparing to that
of non-rotating TNS at the center density of $\rho_{c}=3\rho_{0}$,
where $\rho_{0}$ is the saturation density of nuclear matters.

Fig. 3. Increment of the mass and radius of HS comparing to that
of non-rotating HS at the center density of $\rho_{c}=4\rho_{0}$.

Fig. 4.  The total mass of the non-rotating and rotating TNS as a
function of central density, where the mass is in a unit of solar
masses.

Fig. 5. The radius of the non-rotating and rotating TNS as a
function of central density.

 Fig. 6. The total mass of the non-rotating and rotating HS as a function of central density,
where the mass is in a unit of solar masses

Fig. 7. The radius of the non-rotating and rotating HS as a
function of central density.

Fig. 8. Period as a function of the central density at two
different angular velocities relative to the local inertial frame
at the center.

Fig. 9. The masses as a function of the period at two different
central density for the TNS and HS.

Fig. 10. The radii as a function of the period at two different
central density for the TNS and HS.

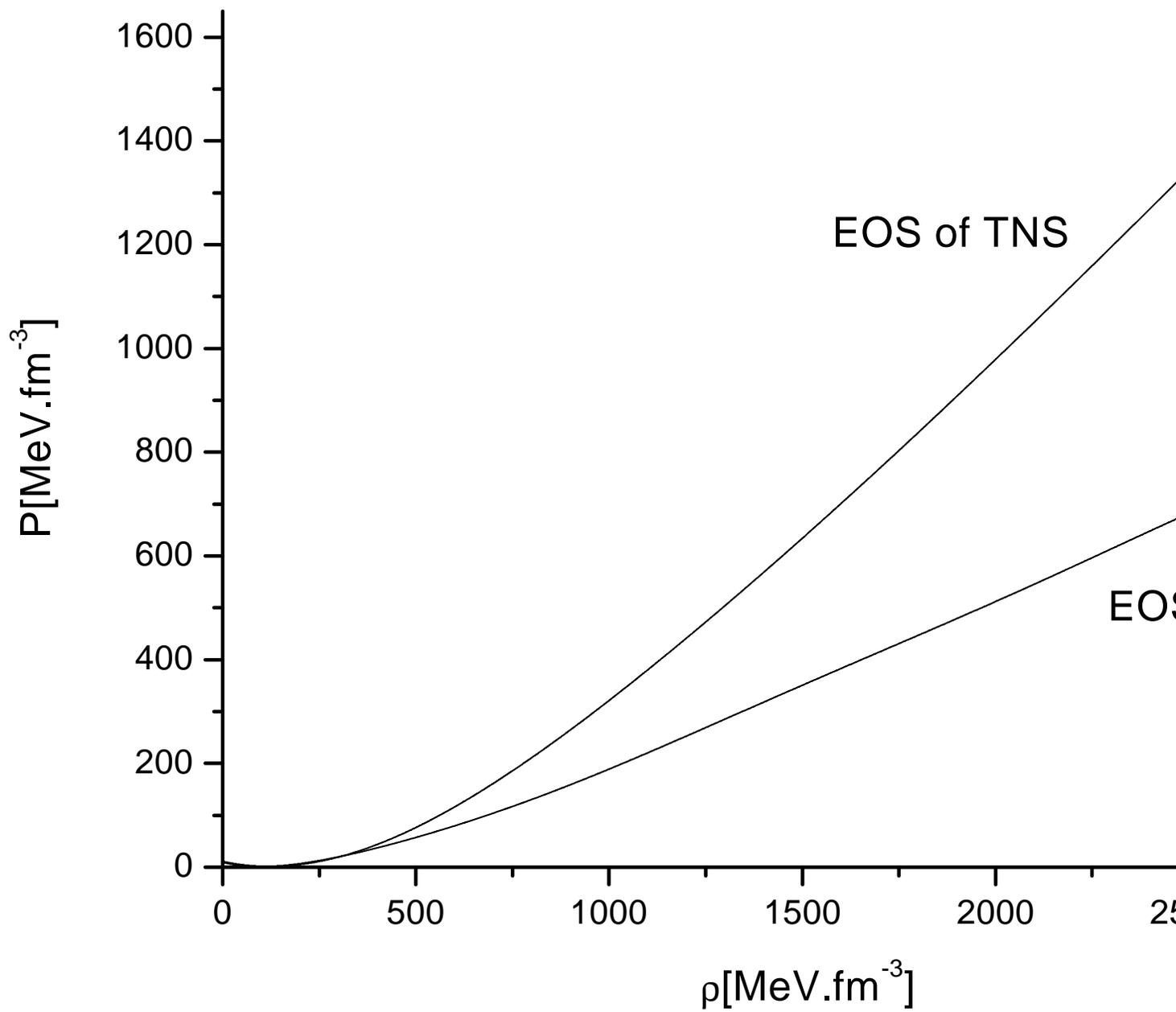
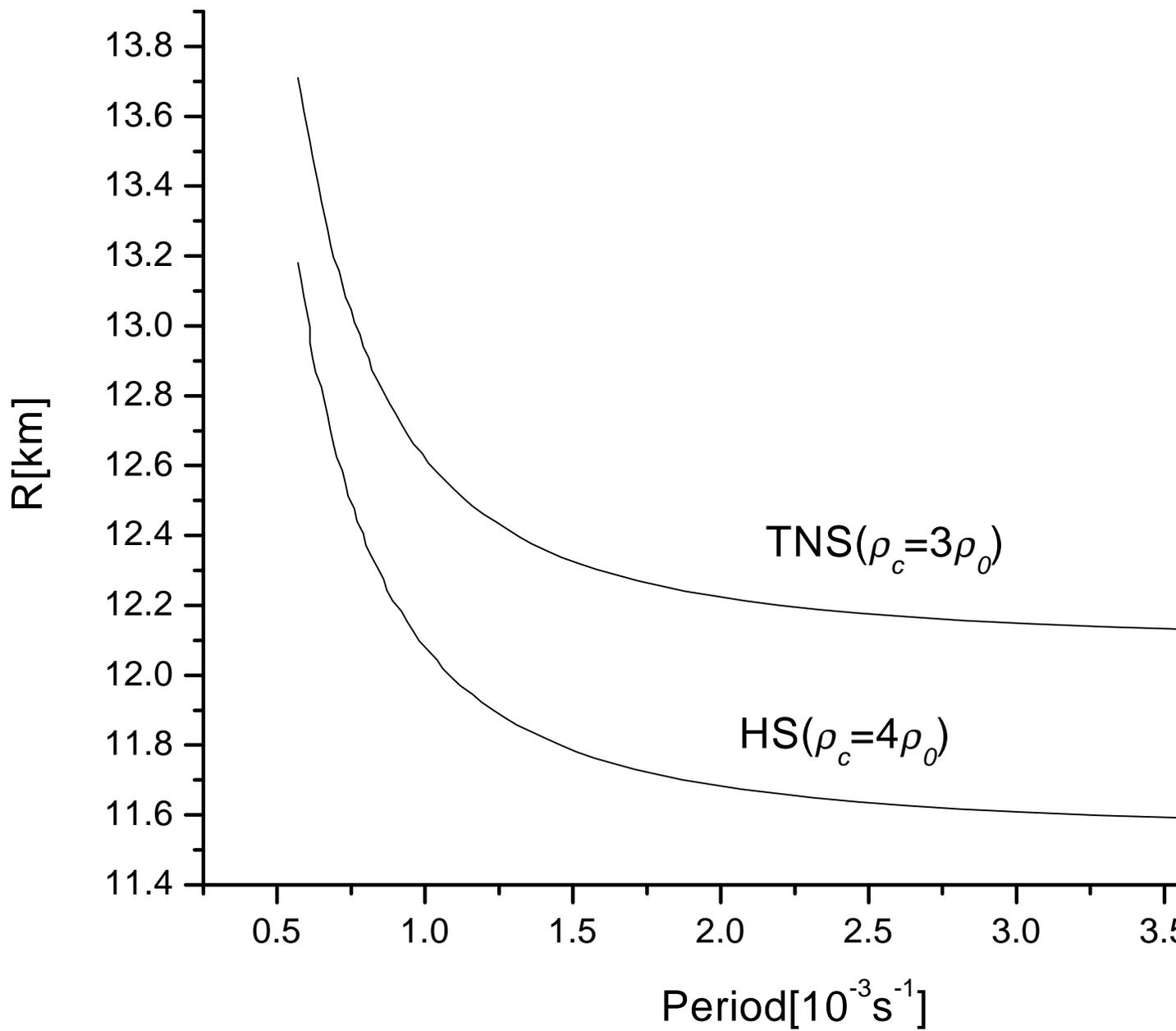
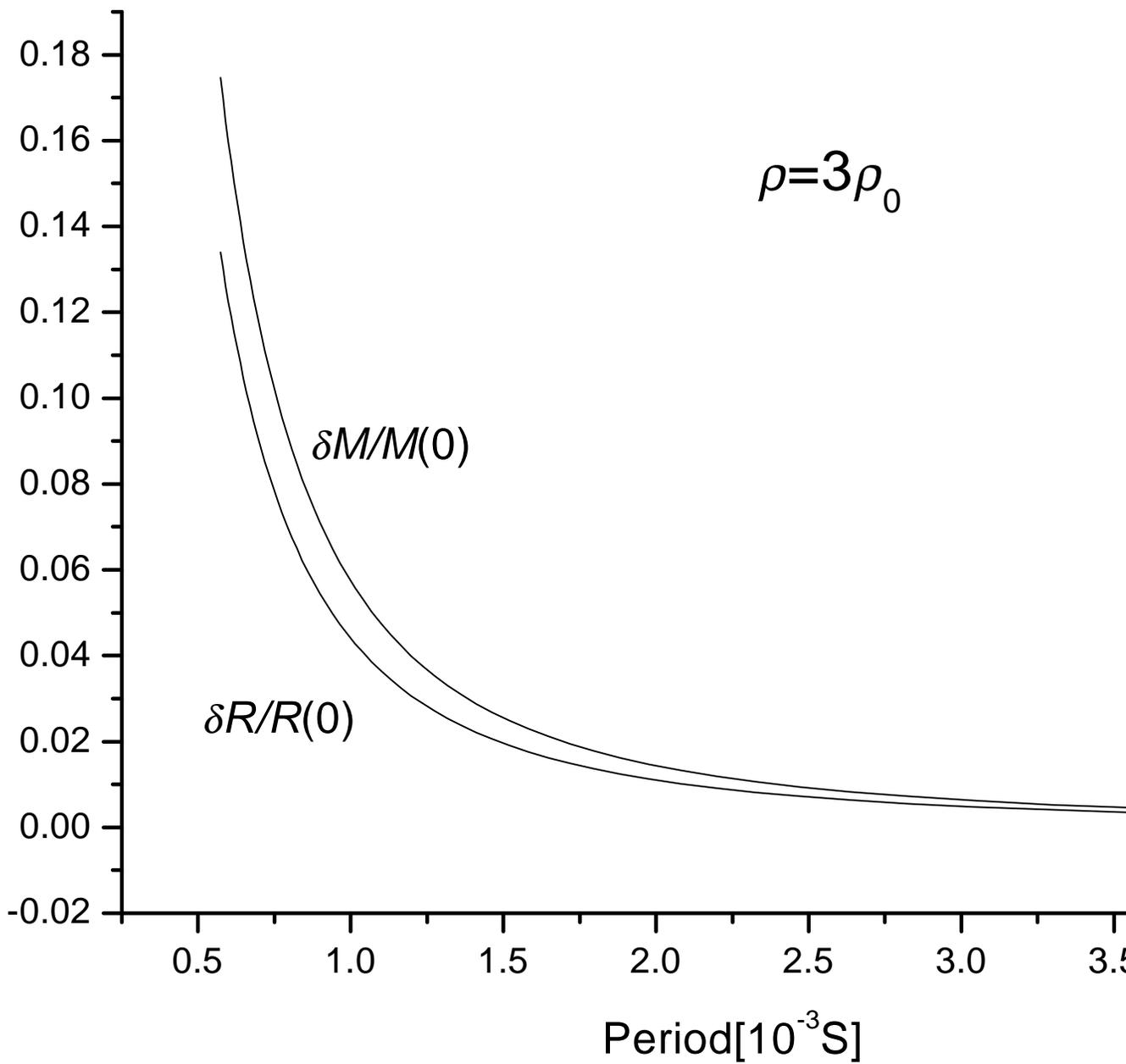
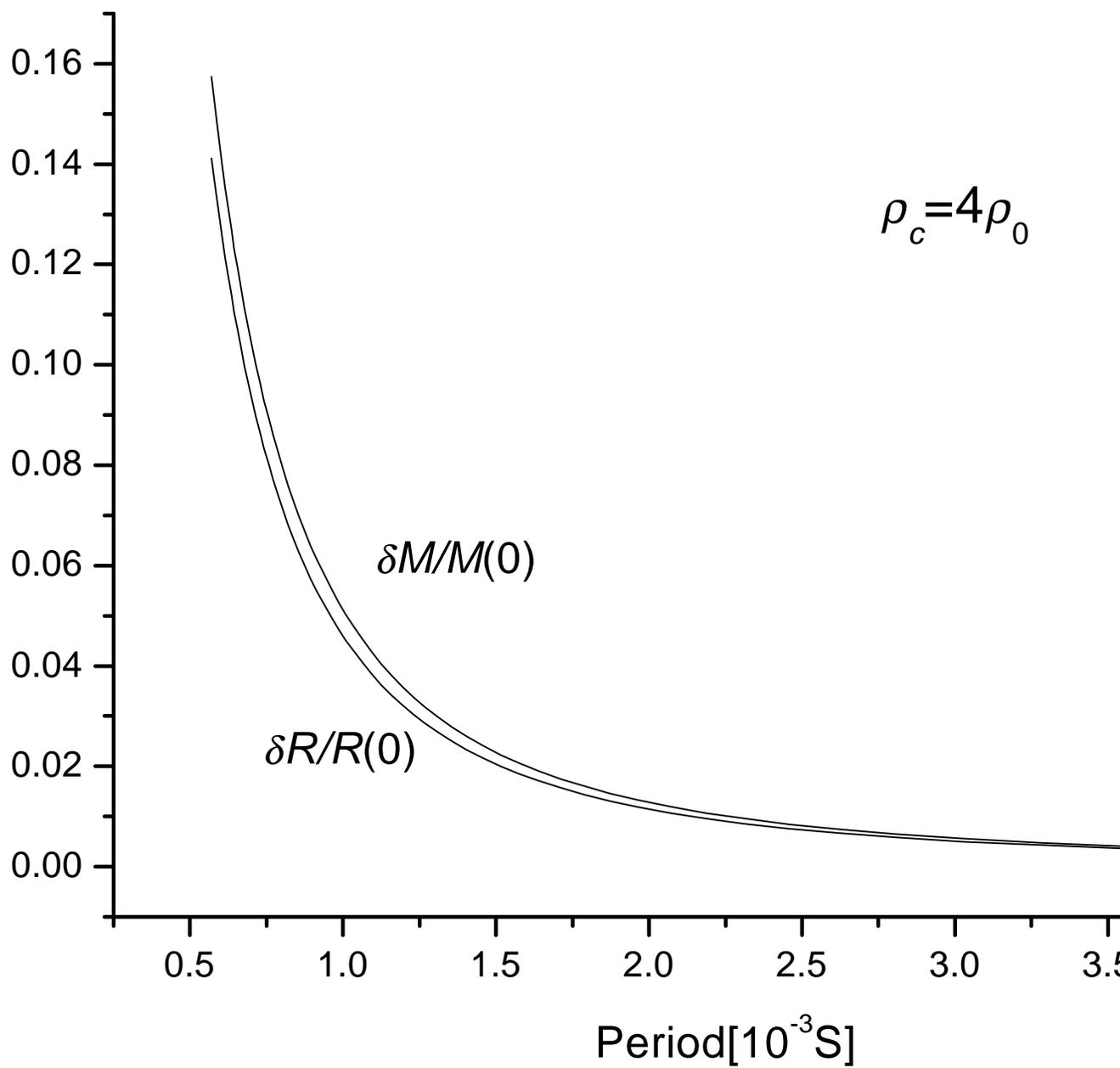
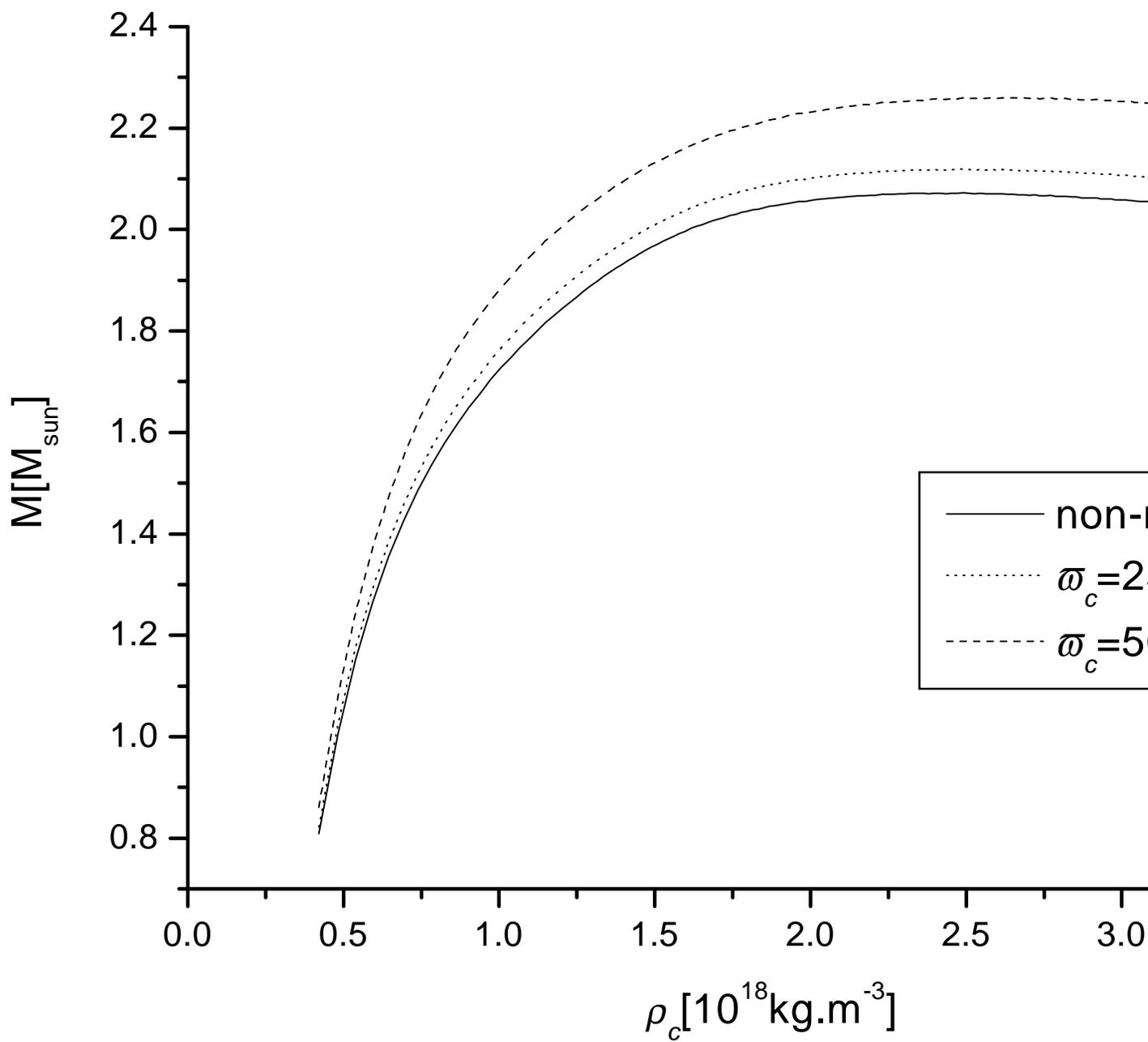
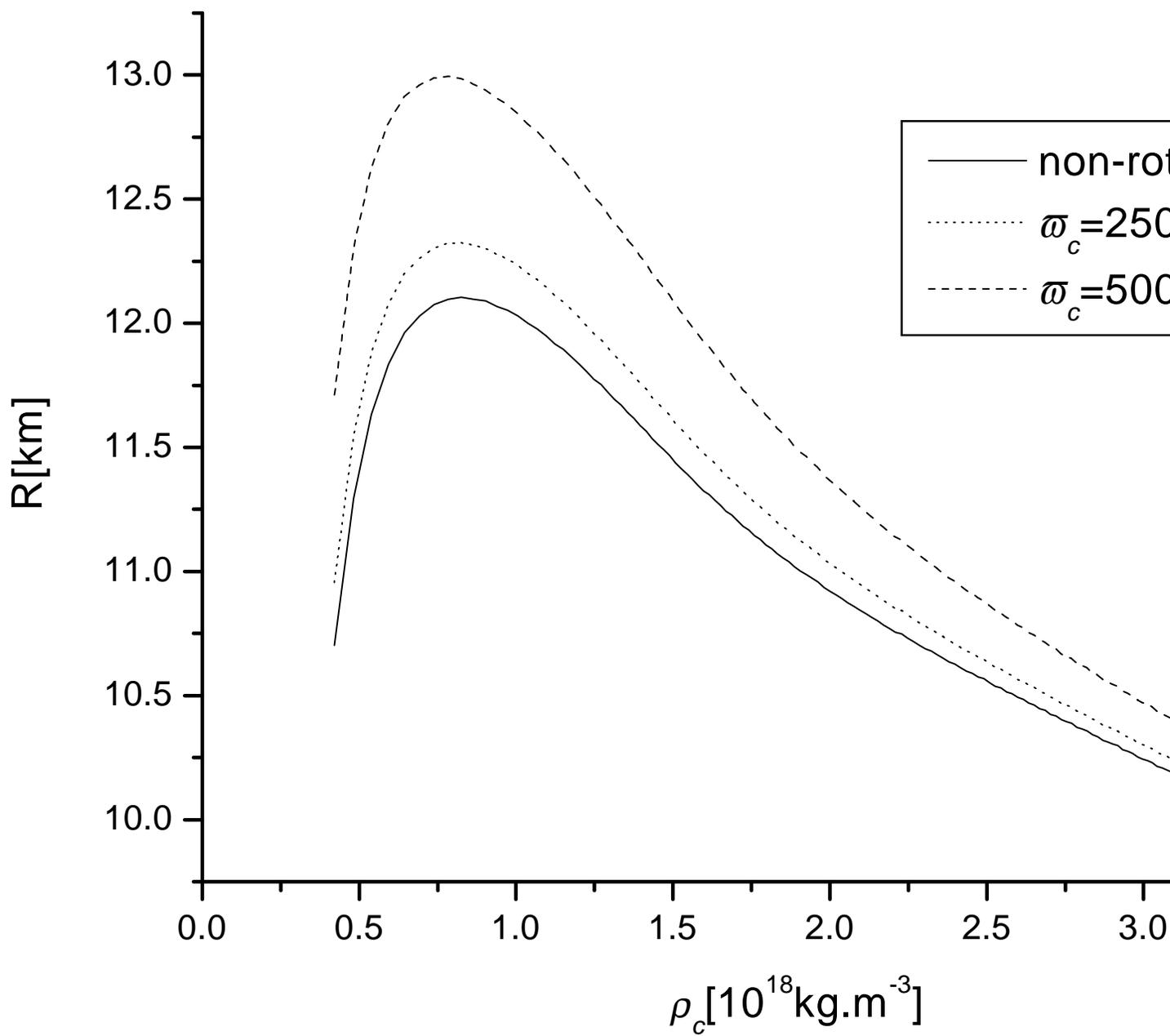
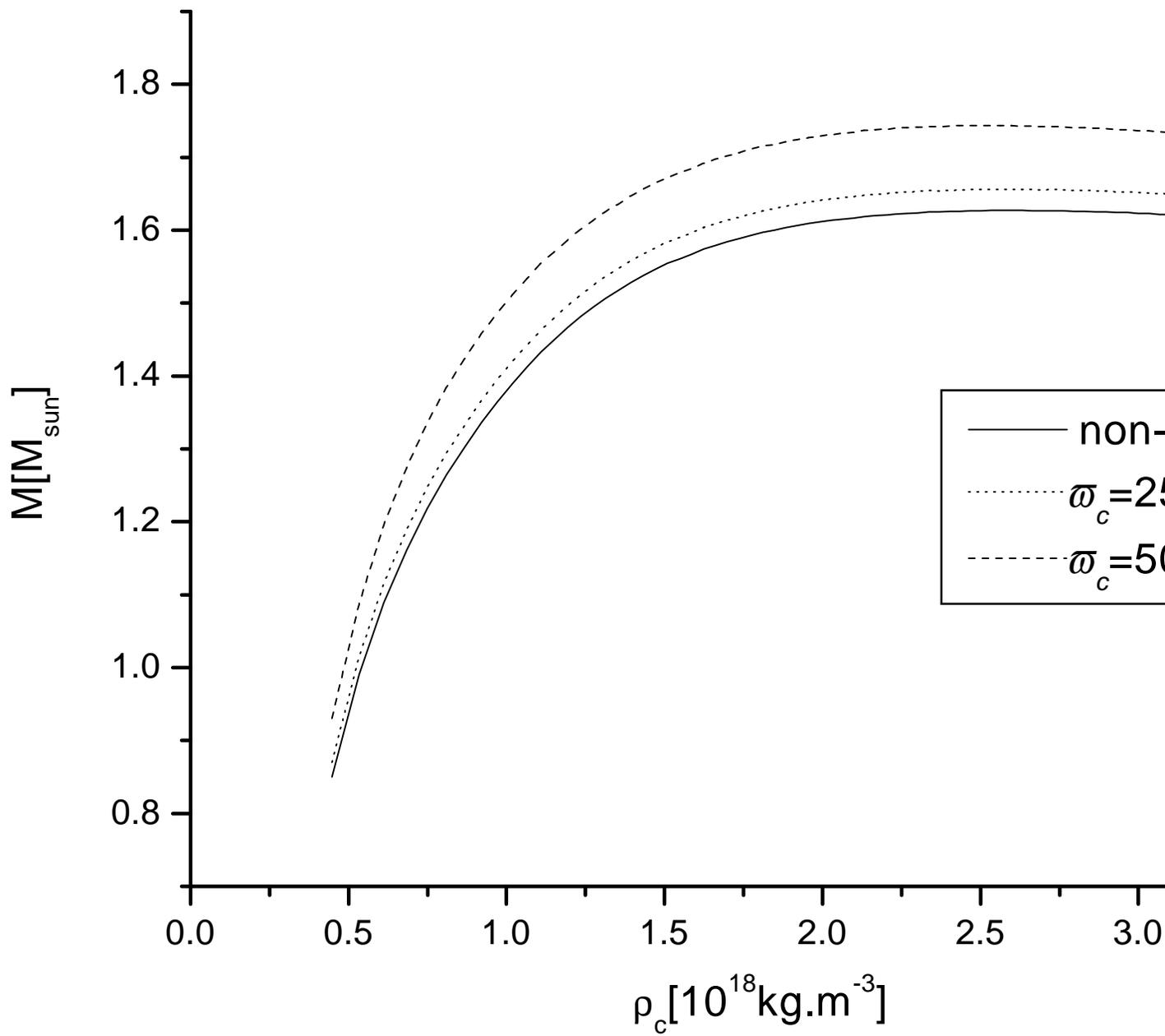
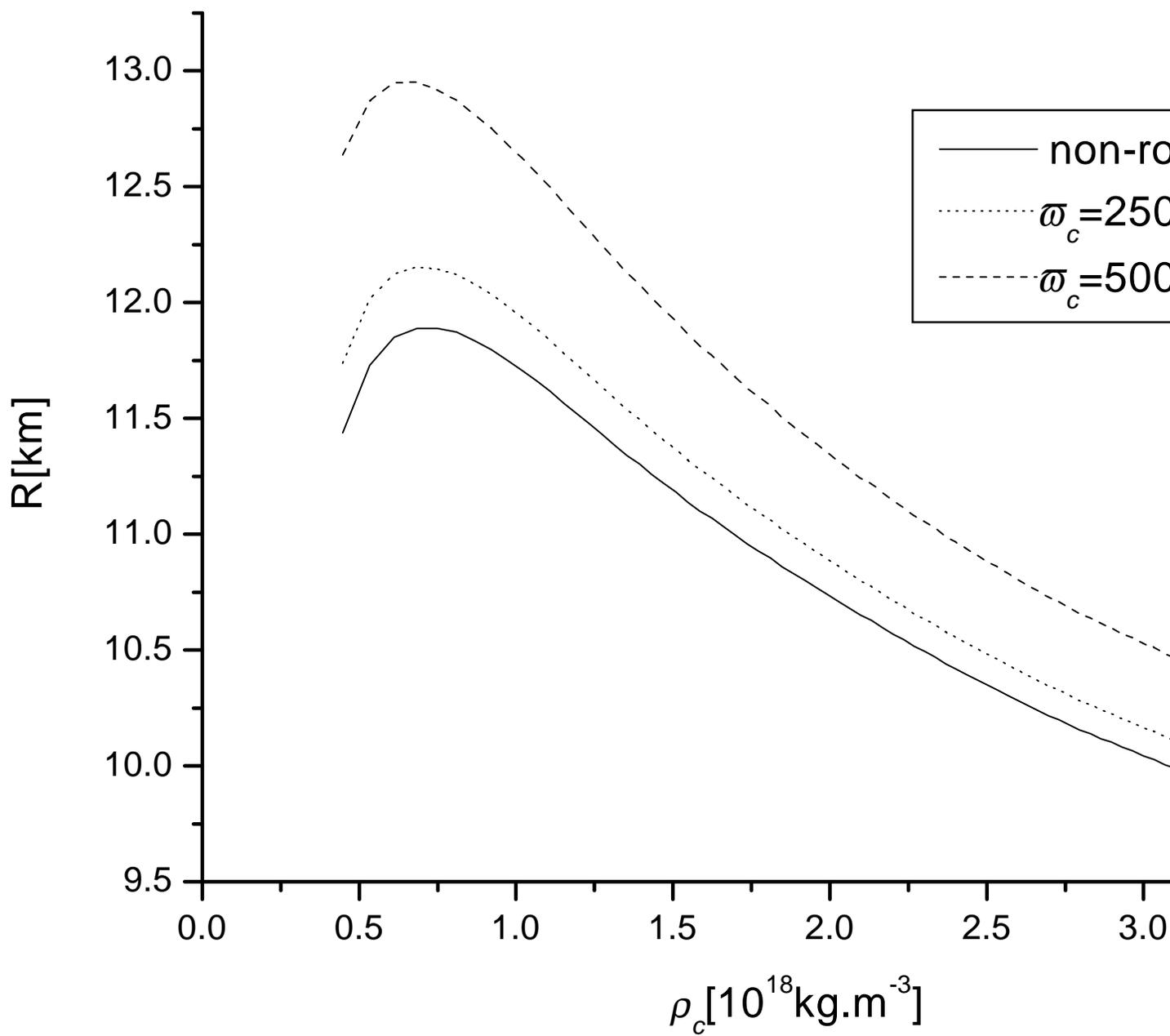
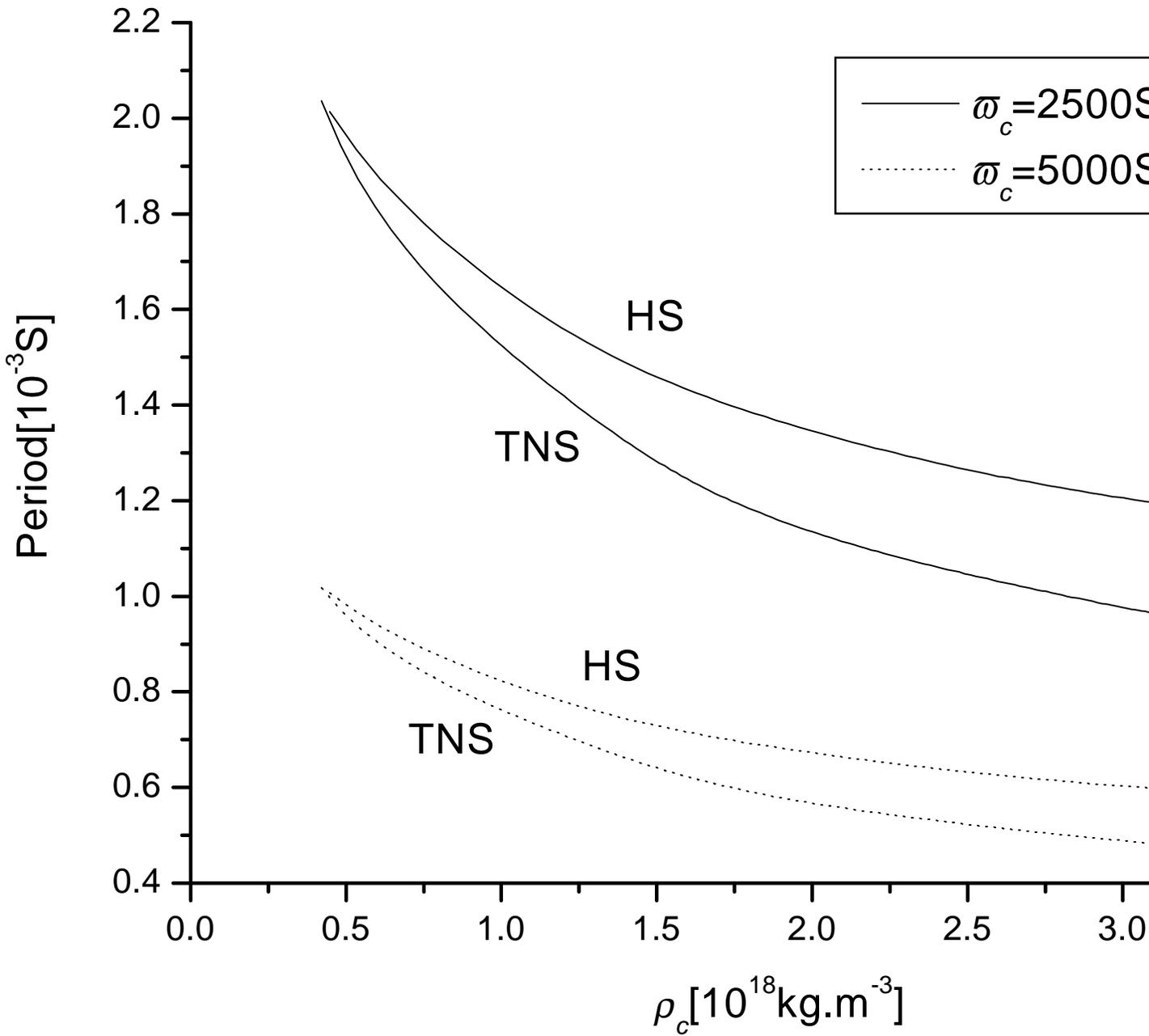
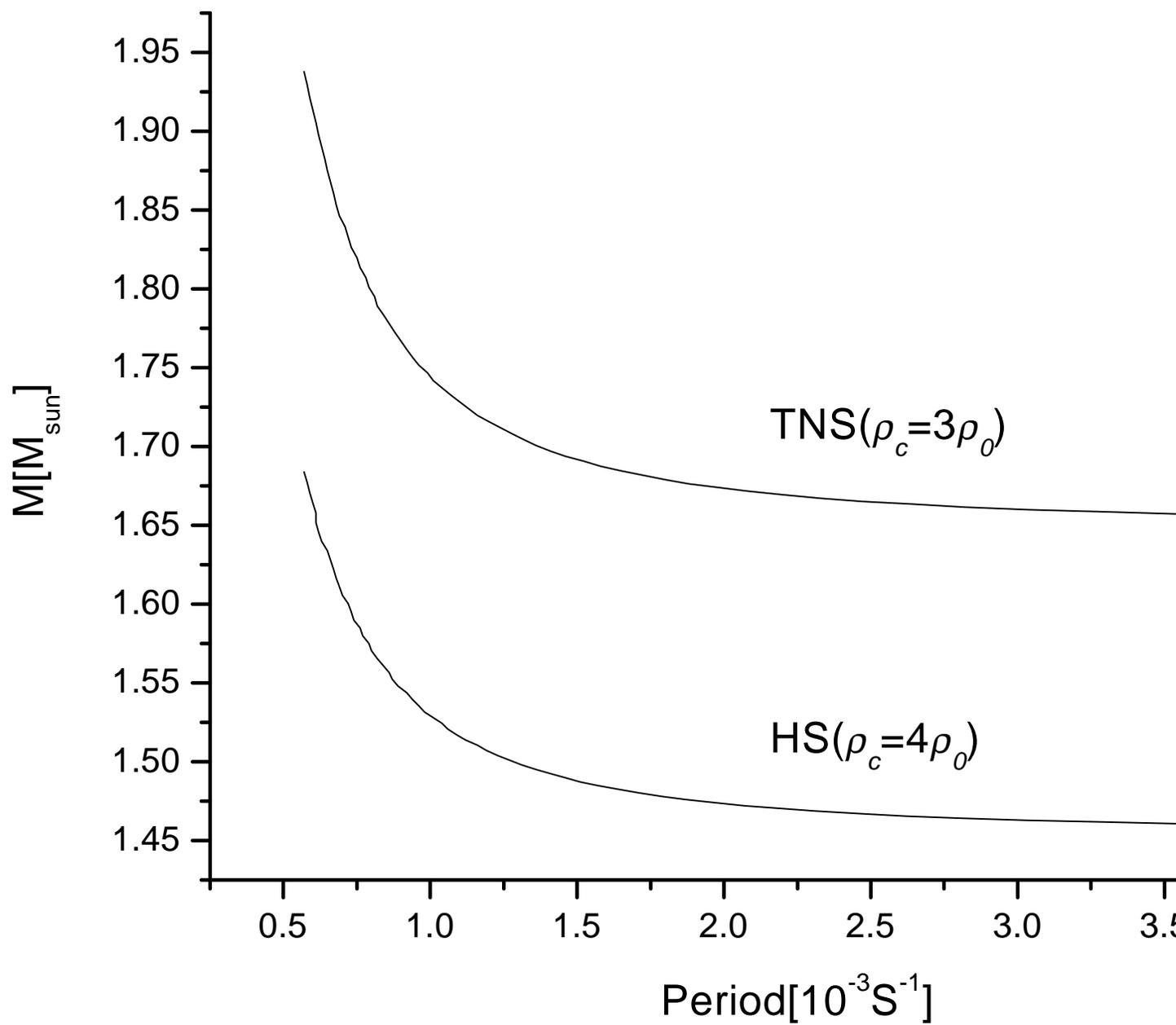
\end{document}